\documentclass[twocolumn,aps,prb,tightenlines,amsmath,amssymb]{revtex4}
\usepackage{graphicx}
\usepackage{amssymb}
\usepackage{dcolumn}
\usepackage{amsmath}
\usepackage{bm}% bold math

\usepackage{colordvi}

\begin{document}

\title{Comment on ``Photon energy and carrier density dependence of
  spin dynamics in bulk CdTe crystal at room temperature''}

\author{J. H. Jiang}
\author{M. W. Wu}
\thanks{Author to whom correspondence should be addressed}
\email{mwwu@ustc.edu.cn.}
\affiliation{Hefei National Laboratory for Physical Sciences at
Microscale and Department of Physics,
University of Science and Technology of China, Hefei,
Anhui, 230026, China}

\date{\today}

\begin{abstract}
We comment on the conclusion by Ma et al. [Appl. Phys. Lett. {\bf 94},
241112 (2009)] that the Elliott-Yafet mechanism is more important
than the D'yakonov-Perel' mechanism at high carrier density in
intrinsic bulk CdTe at room temperature. 
We point out that the spin relaxation is solely
from the D'yakonov-Perel' mechanism. The observed peak in the density
dependence of spin relaxation time is exactly what we predicted in a
recent work [Phys. Rev. B {\bf 79}, 125206 (2009)].
\end{abstract}

\maketitle

In a recent Letter,\cite{Ma} Ma {\em et al}. 
measured the density dependence of electron spin
relaxation time in intrinsic bulk CdTe at room temperature.
They found that the electron spin lifetime first increases then
decreases with increasing excitation density. They attributed the
increase of spin lifetime at low excitation density to the
D'yakonov-Perel' mechanism whereas the decrease at high excitation
density to the Elliott-Yafet mechanism. They concluded that the
Elliott-Yafet mechanism dominates spin relaxation at high excitation
density in CdTe at room temperature.

Their conclusion can not be correct. As shown in our recent
work, the Elliott-Yafet mechanism is unimportant even in narrow band
gap semiconductors such as InAs and InSb for $n$-type and intrinsic
samples in metallic regime.\cite{jiang} For CdTe, which has a large band gap of
$E_g=1.45$~eV, the Elliott-Yafet mechanism can not be important for
intrinsic samples, especially at such high temperature of $300$~K.

Below, through a fully microscopic calculation, we show that the
Elliott-Yafet mechanism is totally {\em irrelevant} to spin relaxation
under the experimental condition. The calculation is based on the
fully microscopic kinetic spin Bloch equation approach with {\em all}
relevant scatterings, such as electron-impurity, electron-phonon,
electron-electron and electron-hole scatterings, explicitly included.\cite{jiang}
The spin-flip process due to the Elliott-Yafet mechanism is {\em
  fully} incorprated in {\em all} these scatterings. The
calculation based on kinetic spin Bloch equation approach has achieved
good agreements with different experiments [e.g., see Appendix~A of
Ref.~\onlinecite{jiang}].

The calculation of the spin relaxation due to the Elliott-Yafet 
mechanism  is based on the following spin-flip scattering,
\begin{equation}
\Gamma_s({\bf k}) = 2\sum_{{\bf k}^{\prime}} \frac{1}{\tau_{p}({\bf
    k}\to{\bf k}^{\prime})} |\Lambda_{{\bf k},{\bf k}^{\prime}}^{\uparrow\downarrow}|^2
\end{equation}
where $\frac{1}{\tau_p({\bf k}\to{\bf k}^{\prime})}$ is the momentum
scattering rate from state ${\bf k}$ to state ${\bf k}^{\prime}$ and
$\Lambda_{{\bf k},{\bf k}^{\prime}}^{\uparrow\downarrow}= -i\lambda_c ({\bf k}\times{\bf
  k}^{\prime})\cdot\mbox{\boldmath$\sigma$\unboldmath}^{\uparrow\downarrow}$.\cite{opt-or}
Here $\lambda_c=\frac{\hbar^2\eta(1-\eta/2)}{3m_cE_g(1-\eta/3)}$ with
$\eta=\frac{\Delta_{\rm SO}}{\Delta_{\rm SO} +
  E_g}$.\cite{opt-or} $m_c$ is the conduction band effective
  mass. $E_g$ and $\Delta_{\rm SO}$ are the band-gap and the
spin-orbit splitting of the valence band, respectively. 
The momentum scattering rate is determined by {\em all} relevant
scatterings, such as the electron-impurity, electron-phonon,
electron-electron and electron-hole scatterings:
$\frac{1}{\tau_p} = \frac{1}{\tau_{\rm ei}}+\frac{1}{\tau_{\rm
    ep}}+\frac{1}{\tau_{\rm ee}}+\frac{1}{\tau_{\rm eh}}$. %The
%electron-impurity scattering rate is $\frac{1}{\tau_{\rm ei}({\bf
%    k}\to{\bf k}^{\prime})}=\frac{2\pi}{\hbar} n_i V_{{\bf k}-{\bf
%    k}^{\prime}}^2 \delta(\varepsilon_{\bf k}-\varepsilon_{{\bf
%    k}^{\prime}})$ where $n_i$ is the impurity density and $V_{{\bf k}-{\bf
%    k}^{\prime}}$ represents the Coulomb
%interaction. $\varepsilon_{\bf k}$ is the electron kinetic energy. The
%electron-phonon scattering rate is $\frac{1}{\tau_{\rm ep}({\bf
%    k}\to{\bf k}^{\prime})}=\sum_{\lambda\pm} \frac{2\pi}{\hbar}
%|M_{\lambda,{\bf k}-{\bf k}^{\prime}}|^2 N^{\pm}_{\lambda,{\bf k}-{\bf k}^{\prime}}
%\delta(\varepsilon_{\bf k}-\varepsilon_{{\bf
%    k}^{\prime}}\mp\hbar\omega_{\lambda,{\bf k}-{\bf k}^{\prime}})$ where
%$\lambda$ is the phonon branch index, $M_{\lambda,{\bf k}-{\bf
%    k}^{\prime}}$ is the electron-phonon interaction matrix element,
% $\omega_{\lambda,{\bf q}}$ stands for the
%phonon dispersion. $N^{\pm}_{\lambda,{\bf q}} = N_{\lambda,{\bf
%    q}}+\frac{1}{2}\pm\frac{1}{2}$ with $N_{\lambda,{\bf
%    q}}=(e^{\hbar\omega_{\lambda,{\bf q}}/k_BT}-1)^{-1}$ being the
%phonon number. The momentum scattering rate due to the
%electron-electron and electron-hole scatterings, which could be found
%in Ref.~\onlinecite{jiang}, will not be presented in detail here.
The corresponding scatterings are given in detail in Ref.~\onlinecite{jiang}.

The spin relaxation time $\tau_s$ is then obtained by average
over $\Gamma_s({\bf k})$, $1/\tau_s=\langle\Gamma_s({\bf k})\rangle$. 
It is noted that there is {\em no fitting
parameter} in the calculation. %The matrix element of
%electron-phonon interaction is taken from textbooks and the Coulomb
%interaction is treated within the standard random phase
%approximation. 
The material parameters of CdTe are taken from the
standard handbooks of {\em Landolt-B\"ornstein}.\cite{para}

From the parameter-free fully microscopic calculation, we obtain the
spin lifetime limited by the Elliott-Yafet mechanism, $\tau_s\gtrsim
800$~ps in the excitation density range of $10^{14}$ to
$10^{17}$~cm$^{-3}$. This is at least two-orders of magnitude larger than
the one observed in the experiment by Ma {\em et al}.\cite{Ma} Hence the
Elliott-Yafet mechanism is totally {\em irrelevant} under the
experimental condition. The spin relaxation is then solely determined
by the D'yakonov-Perel' mechanism.

As we have pointed in a recent work,\cite{jiang} the density
dependence of spin relaxation time due to the D'yakonov-Perel'
mechanism $\tau_s\sim 1/[\langle \Omega({\bf k})^2\rangle\tau_p]$ is
nonmononotic in intrinsic bulk III-V semiconductors: spin relaxation
time increases with increasing density in non-degenerate regime due to
decrease of momentum scattering time $\tau_p$ but 
decreases in degenerate regime due to the enhancement of inhomoegeneous
broadening $\langle \Omega({\bf k})^2\rangle$. There is a peak in the
crossover regime. For II-VI semiconductors with zinc-blende structure,
the spin-orbit coupling and the band structure is similar to III-V
semiconductors. Hence the same behavior is also expected. Actually, the
band and material parameters of CdTe are very similar to GaAs. In
intrinsic GaAs at room temperature, the peak density is $9\times
10^{16}$~cm$^{-3}$.\cite{jiang} In the experiment by Ma {\em et al}., the
peak density is $3\times 10^{11}$~cm$^{-2}$. As the authors did not
determine the penetration depth of the laser, a rough estimation
gives the peak density of $6\times 10^{16}$~cm$^{-3}$ which is close
to the one in GaAs. This indicates that the observed peak in density
dependence of spin lifetime should be samilar to what we have
predicted in III-V semiconductors.\cite{jiang} However, due to the
uncertainty in the penetration depth and the possible effect of 
hot-electron effect of
the photo-excited carriers (as also indicated by the experimental results in the
photon energy dependence), it is premature  to give a quantitative
comparison.

%In conclusion, the Elliott-Yafet mechanism is totally irrelevant for spin
%relaxation in the experiment by Ma {\em et al}. The observed peak in the
%density dependence of spin relaxation time should be what we predicted
%in a recent work.\cite{jiang}

This work was supported by the Natural Science Foundation of China
under Grant No.~10725417.


\begin{thebibliography}{0}
\bibitem{Ma} H. Ma, Z. Jin, G. Ma, W. Liu, and S. H. Tang,
  Appl. Phys. Lett. {\bf 94}, 241112 (2009).
\bibitem{jiang} J. H. Jiang and M. W. Wu, Phys. Rev. B {\bf 79},
  125206 (2009).
\Red{%\bibitem{note} Assuming a penetration depth of 100~nm, we converted the
%  two-dimensional excitation density in the experiment into three
%  dimensional one.
}
\bibitem{opt-or} F. Meier and B. P. Zakharchenya, {\em Optical
    Orientation} (North-Holland, Amsterdam, 1984).
\Green{%\bibitem{spintronics} I. \v Zuti\'c, J. Fabian, and S. Das Sarma,
  %Rev. Mod. Phys. {\bf 76}, 323 (2004); J. Fabian, A. Matos-Abiague,
  %C. Ertler, P. Stano, and I. \v Zuti\'c, Acta Phys. Slov. {\bf 57},
  %565 (2007).
}
\bibitem{para} {\it Semiconductors}, Landolt-B\"ornstein, New Series,
  Vol.~17b, ed. by O. Madelung (Springer-Verlag, Berlin, 1987).

\end{thebibliography}
\end{document}